\documentclass[aps,prl,twocolumn,showpacs,superscriptaddress]{revtex4-1}
	
	\usepackage{graphicx}
	\usepackage{amsfonts, amsmath, amstext, amssymb, amsfonts, amsxtra}
	\usepackage{braket}
	\usepackage{bbm}
	\usepackage{bbold}
	\usepackage[protrusion=true,expansion=true]{microtype}
	\usepackage{wasysym}
	\usepackage{cleveref}
	\usepackage{color}
	\usepackage{ulem}
	
	\newcommand{\im}{{i}}         
	
	
	
\begin{document}
		
		\title{Photon waiting time distributions: a keyhole into dissipative quantum chaos}
		\author{I.I.~Yusipov$^1$, O.S.~Vershinina $^1$, S.V.~Denisov$^{2,3}$, and M.V.~Ivanchenko $^{1,3}$}
		
		\affiliation{ $^1$ Department of Applied Mathematics, Lobachevsky University, Nizhny Novgorod, Russia \\
			$^2$ Department of Computer Science, Oslo Metropolitan University, Oslo, Norway\\
		$^3$ Max Planck Institute for the Physics of Complex Systems, Dresden, Germany}
		
		\begin{abstract}

		Open quantum systems can exhibit complex states, which classification and quantification is still not well resolved. The Kerr-nonlinear cavity, periodically modulated in time by coherent pumping of the intra-cavity photonic mode, is one of the examples. Unraveling  the corresponding Markovian master equation into an ensemble of quantum trajectories and employing the recently proposed calculation of quantum Lyapunov exponents [I.I. Yusipov {\it et al.}, Chaos {\bf 29}, 063130 (2019)], we identify `chaotic' and `regular' regimes there. In particular, we show that chaotic regimes manifest an intermediate power-law asymptotics in the distribution of photon waiting times. This distribution can be retrieved by monitoring photon emission with a single-photon detector, so that  chaotic and regular states can be discriminated without disturbing the intra-cavity dynamics.
		      
		\end{abstract}

		\maketitle

		\textbf{Quantification of regimes emerging in open quantum systems  driven out of equilibrium is a problem, interesting in several respects. In particular, it could help bridging non-equilibrium  quantum phenomena with manifestations of classical dissipative 
		chaos, such as local instability, bifurcations, strange attractors, etc \cite{Ott}.	Considerable effort was devoted to quantum generalization of Lyapunov exponents (LEs), the most popular and powerful means to quantify classical  chaos \cite{Toda1987,Karol1992,Manko2000}. Recently, a  new wave of research in this direction has been initiated by the idea of out-of-time  correlation functions as possible quantum LE analogs	\cite{Galitski2017,Galitski2018,Hirsch2019}. In our recent work \cite{Yusipov2019}, we proposed yet another approach to `quantization' of LEs, based on unraveling of the Markovian 
		master equation that describes the evolution of the system's density matrix into an ensemble of trajectories. 
		However, all the approaches suffer from a common problem: While it is possible to define the corresponding observables and correlators and then treat them 
		analytically or calculate numerically, it is much harder (or not possible at all)  to access them in a real-life experiment. Here we demonstrate that 
		`chaos-regularity' transitions can be captured by looking at the statistics  of dissipative jump events that are  accessible 
		in quantum optics and called there ``distribution  of photon waiting times''~\cite{Carmichael1991,Flindt2019}.}

		\textbf{Exemplifying in a simple model of an open periodically modulated quantum system,	we show that transitions from regular to chaotic regimes (beforehand defined in terms of LEs) are marked by transitions from exponential waiting-time distributions to those with intermediate power-law scaling. Since photon emission events can be resolved in an experiment, e.g. by using single-photon detectors, discrimination between chaotic and regular regimes can therefore be achieved without disturbance of the system dynamics. 
		}

		\section{Introduction}
		
		A recent progress in such fields of  experimental quantum physics  as cavity quantum electrodynamics \cite{Walter2006}, cavity optomechanics \cite{Aspelmeyer2014}, 
		artificial atoms \cite{You2011} and polaritonic devices \cite{Feurer2003}, gave the theory of open quantum systems \cite{book} a new impetus. Different from the typical single-particle 
		models of quantum optics \cite{Carmichael1991}, such systems are essentially multi-component and thus require many-body models to  describe them.
		These models, when considered out of equilibrium,  display asymptotic states that are sometimes interpreted as quantum versions of 
		chaotic (or regular) attractors \cite{Ott}. Until recently, such classification was  performed mostly visually, after projecting quantum states onto a suitable classical phase space \cite{Spiller1994,Brun1996,Hartmann2017,Ivanchenko2017,Poletti2017,Yusipov2019}. 
		The quantitative approach to \textit{dissipative} Quantum Chaos  \cite{Haake2010} is possible by means of quantum versions of Lyapunov exponents (LEs) \cite{Pikovsky}. 
		There is no unique definition of quantum LEs; 	instead, there are several  alternative generalizations  ranging from early attempts to define the exponents in terms 
		of quasi-classical distributions \cite{Toda1987,Karol1992,Manko2000} to stochastic Schr\"{o}dinger equation \cite{Persival1992,Jacobs2000,Pattanayak2008,Jacobs2009,Pattanayak2009,Pattanayak2018}
		and further on, to a very recent approach based on the concept of out-of-time-correlators \cite{Galitski2017,Bohrdt2017,Galitski2018,Hirsch2019}. 
		However, all these generalizations suffer a common problem: the corresponding quantifiers are hard (or not possible at all) to measure in a real-life  experiment.

		In our recent work \cite{Yusipov2019}, we introduced a version of quantum LEs based on the so-called  `quantum trajectory' unraveling \cite{plenio,daley} 
		(also called `Monte Carlo wave-function (MCwf) method' \cite{zoller,dali}), 
		which replaces the original  quantum Markovian master equation \cite{book} with an ensemble of quantum trajectories. The evolution of the trajectories is governed by a non-Hermitian Hamiltonian in a continuous-time manner, interrupted by random jump-like dissipative events. 	
		While this version, in principle, opens a way to theoretical characterization of quantum dissipative chaos in physically relevant setups (e.g., in  cavity quantum electrodynamic systems \cite{Walter2006}), the experimental estimation of such quantum LEs is no less challenging than for the prior versions. 
		
		Here we propose an experimentally feasible approach to detecting chaotic regimes in cavity-like open quantum systems. Namely, using 
		a periodically modulated Kerr-nonlinear cavity as a model, we demonstrate that the transition to quantum chaos, quantified with  the LEs from Ref. \cite{Yusipov2019},
		is associated with the appearance of an intermediate power-law asymptotics 
		in the distribution of waiting time (that is the time between two consequent photon emissions from the cavity \cite{Vyas,Carmichael1989,baran}). 
		This distribution can be sampled in an experiment by using single-photon detection techniques \cite{photon1,photon2}.

		\section{Model}  We consider a photonic mode in a leaky Kerr-nonlinear cavity, periodically modulated 
		by an external coherent EM field \cite{Spiller1994,Brun1996}. Its unitary dynamics is governed by the Hamiltonian (we set $\hbar = 1$)
		\begin{equation}
		H(t) = \frac{1}{2}\chi \hat{a}^{\dagger 2}\hat{a}^2 +iF(t)(\hat{a}^{\dagger} - \hat{a}).
		\label{eq:2}
		\end{equation}
		Here, $\chi$ is the photon interaction strength, $\hat{a}^\dagger$ and $\hat{a}$ are 
		photon creation and annihilation operators, so that $\hat{n}=\hat{a}^{\dagger}\hat{a}$ is the photon number operator. $F(t)=F(t+T)$ describes periodic modulation. 
		We use the two-valued quench-like driving with period $T$, that is $F(t) = A$ within $0 < t \le T/2$ and $F(t) = 0$ for the 
		second half period $ T/2 < t \leq T$.

		Photons can be emitted from the cavity. In principle, 
		they can also be pumped in by thermal environment, but here, similar to the setups in Refs. \cite{Spiller1994,Brun1996},
		we work in the zero-temperature limit, assuming that the pumping rate is zero. Evolution of the cavity is described by the Lindblad master equation \cite{book, alicki},
		\begin{align}
			\dot{\varrho} = \mathcal{L}(\varrho) = -\im [H,\varrho] + \mathcal{D}(\varrho),
			\label{eq:1}
		\end{align}
		where the first term in the r.h.s.\ captures the unitary dynamics of the system, determined by Hamiltonian (1), while 
		the second term reflects coupling to the environment. The dissipative term involves a single jump operator,
		\begin{equation}
		\begin{aligned}
			\mathcal{D}(\varrho) &= V\varrho V^\dagger - \frac{1}{2}\{V^\dagger V,\varrho\}, ~~V&=\sqrt{\gamma}\hat{a}, 
		\end{aligned}
		\label{eq:3}
		\end{equation}
		which describes  emission of photons by the cavity into the zero-temperature environment. The
		dissipative coupling constant $\gamma$ is assumed to be time-independent.  
		
		In simulations, we limit the number of photons in the cavity mode to $N$, so that the Hilbert space of the numerical model has dimension $N+1$ and can be spanned  with the $N+1$ Fock basis vectors, $\{|n+1\rangle\}$, $n=0,...,N$. $N$ is chosen to be large enough so that the average number of photons in the cavity, 
		$\langle N_{\mathrm{ph}} \rangle$, remains substantially smaller. $\langle N_{\mathrm{ph}} \rangle$ depends on parameters of Hamiltonian; yet the main control
		parameter, which allows to control the mean number of photons is the coupling $\chi$ \cite{Spiller1994,Brun1996}.
		Throughout the paper we set $\chi=0.008$ and $\gamma=0.05$. It gives $\langle N_{\mathrm{ph}} \rangle \sim 50$ so we set $N =300$ and verify that it suffices.

		We make use of the quantum Monte-Carlo wave function method 
		to unravel the deterministic  equation (\ref{eq:1}) into an ensemble of quantum trajectories \cite{zoller,dali,plenio,daley}. 
		It allows us to describe evolution of the model system in terms of an ensemble of pure states, $\psi(t)$, governed by the effective non-Hermitian Hamiltonian \cite{Spiller1994,Brun1996},
		\begin{align}
			i\dot{\psi}=H\psi -\frac{i}{2} V^\dagger V\psi.
			\label{eq:4}
		\end{align} 
		The norm of the wave function decays according to 
		\begin{align}
		\frac{d}{dt}||\psi||= -\psi^* V^\dagger V\psi,
		\label{eq:4a}
		\end{align} 
		until it reaches a threshold $\eta$, repeatedly chosen as i.i.d. random number from $[0,1]$. Then a random jump is performed, and the wave function norm is reset to $||\psi(t)||=1$. 
		After that the continuous non-unitary evolution,  Eq.~(\ref{eq:4}), is resumed until the next quantum jump, etc. For the model given by  Eqs.(\ref{eq:2},\ref{eq:3}), 
		a quantum jump corresponds to emission of a single photon, which can be recorded by a photodetector \cite{Carmichael1991}.

		The density matrix sampled from a set of $M_r$ realizations, that originate from an initial pure state $\psi^\mathrm{init}$ for Eq.~(\ref{eq:4}), as $\varrho(t_\mathrm{p};M_{\mathrm{r}}) = \frac{1}{M_{r}}
		\sum_{j=1}^{M_{\mathrm{r}}} \ket{\psi_j(t_\mathrm{p})}\bra{\psi_j(t_\mathrm{p})}$,
		converges towards the solution of Eq.~(\ref{eq:1}) at time $t_\mathrm{p}$ for the initial density matrix $\varrho^{\mathrm{init}} = \ket{\psi^\mathrm{init}}\bra{\psi^\mathrm{init}}$.

		\begin{figure}[t]
			\begin{center}
				(a) \includegraphics[width=0.9\columnwidth,keepaspectratio,clip]{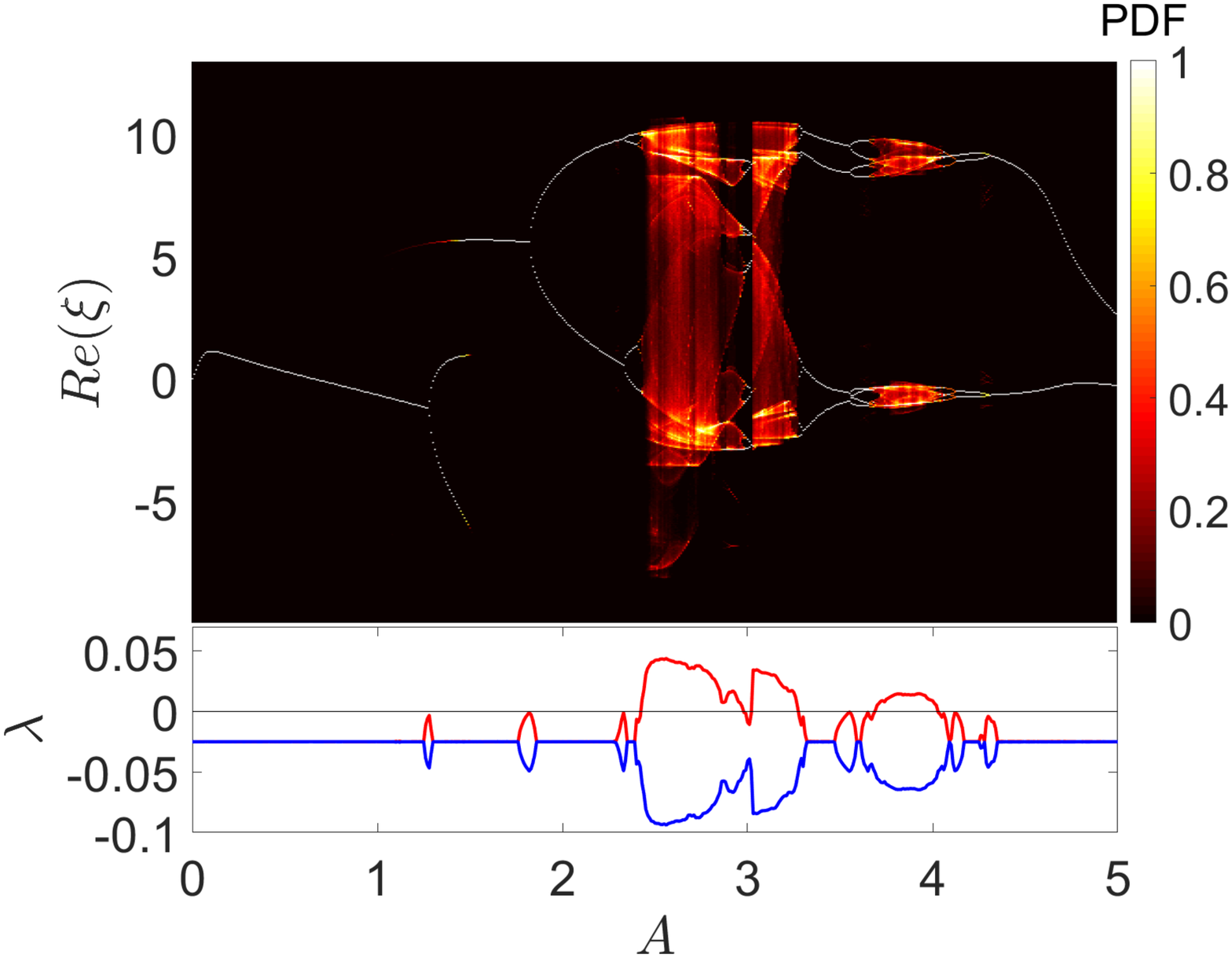}\\
				(b) \includegraphics[width=0.9\columnwidth,keepaspectratio,clip]{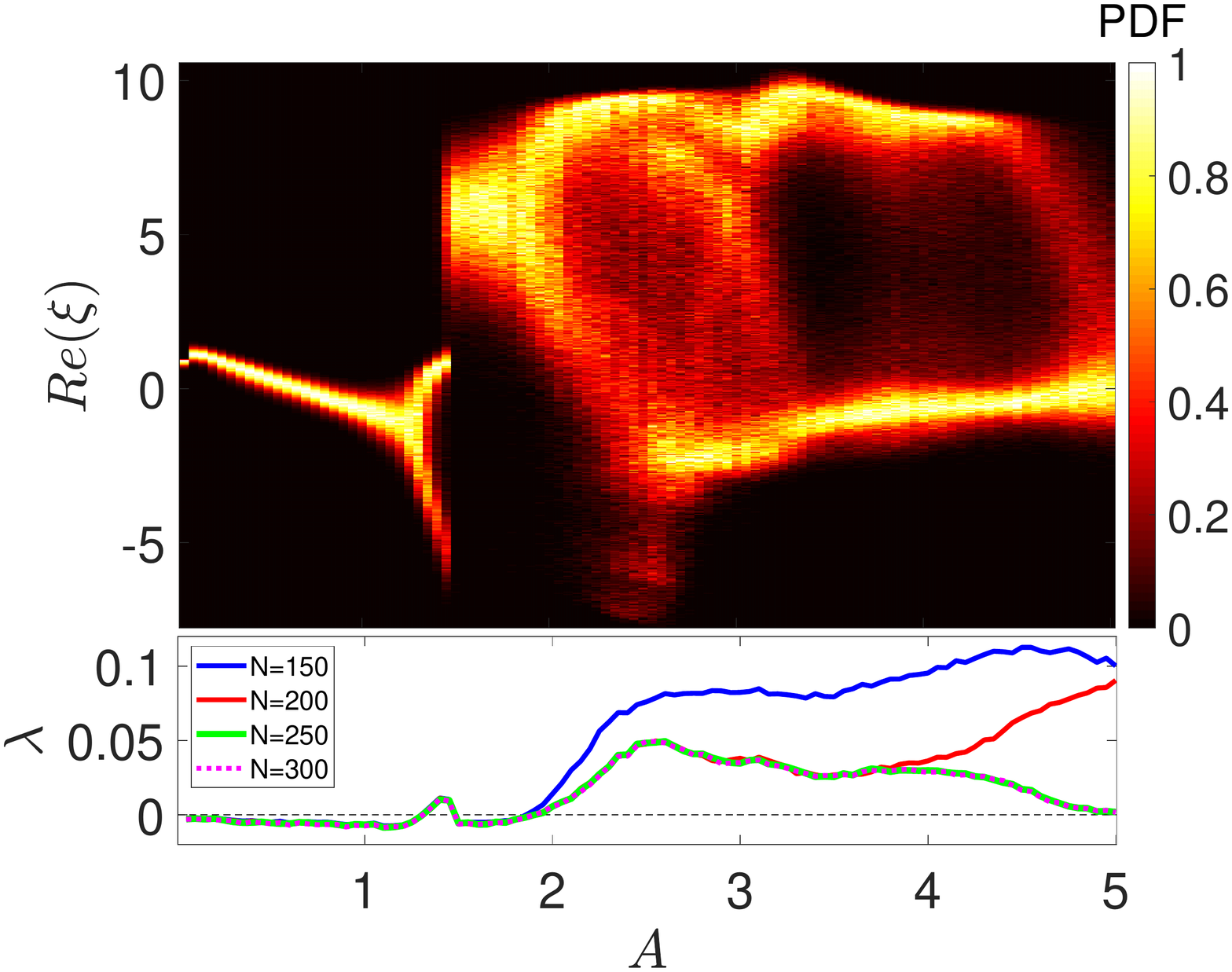}
				\caption{(Color online) Bifurcation diagrams and largest Lyapunov exponent for the model system as functions of 
				the modulation amplitude $A$ in the classical (a) and quantum (b) cases. The  color-coded probability distributions for $Re(\xi)$ are normalized so that the maximum for every $A$ value is set to $1$. The other parameters are $T=10, N=300$.  The values of quantum LE are consistenly reprodiced independent on the dimension of the numerical model for $N\ge250$, (b) bottom panel.} 
				\label{fig:3}
			\end{center}
		\end{figure}

		Following Refs.~\cite{Spiller1994,Brun1996}, we use the complex valued observable of the non-Hermitian photon annihilation operator as a dynamical `variable'  
		\begin{align}
			\xi(t)&=\langle\psi^\dagger(t)|\hat{a}|\psi(t)\rangle.
			\label{eq:5}
		\end{align}
		In the mean-field classical limit, $N_{\mathrm{ph}} \rightarrow \infty$,  its evolution is given by 
		\begin{eqnarray}
		\dot{\xi} = -\frac{1}{2}\gamma\xi + F(t) - i\chi|\xi|^2\xi.
		\label{eq:7}
		\end{eqnarray}
		
		The state of the mean-field system is described by two phase variables, $\{\mathrm{Re}~\xi, \mathrm{Im} ~\xi \}$. This system is essentially non-linear and periodically modulated time
		and, expectedly, exhibits a spectrum of different asymptotic regimes, from varios periodic orbits to chaotic attractors. The parameter dependence can be visualized with a bifurcation diagram, Fig.~\ref{fig:3}(a), constructed from stroboscopic values, $\xi_k=\xi(t_0+kT)$, where $t_0$ is a transient time given to the system to land on the corresponding attractor. 
		In particular, one observes that the largest Lyapunov exponent becomes positive upon transition from a fixed point to a chaotic attractor in the stroboscopic map (see  Fig.~\ref{fig:3}b).

		\begin{figure}[t]
			\begin{center}
				\includegraphics[width=0.95\columnwidth,keepaspectratio,clip]{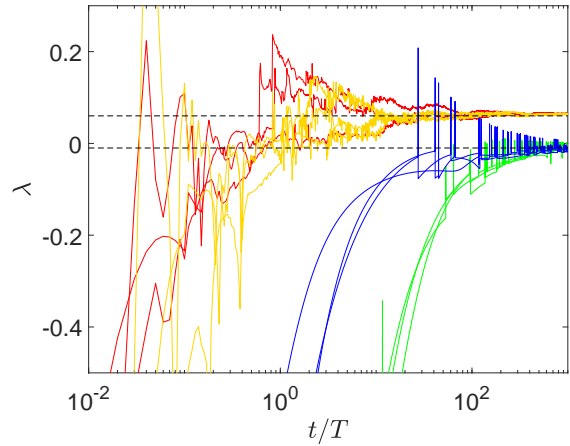}
				\caption{(Color online) Convergence of finite time LEs to their asymptotic values,
				$\lambda=0$ (regular dynamics)  and $\lambda\approx0.08$ (chaotic regime). Three individual trajectories are used in each case. 
				Exponents are calculated by using, first, $\xi(t)$ as an observable, $A=0.05, T=0.5$ (blue), $A=4.0, T=20$ (red), and, second, the mean number of photons in the cavity, $n(t)$, $A=0.05, T=0.5$ (green), $A=4.0, T=20$ (orange). Here $N=300$.}  
				\label{fig:2}
			\end{center}
		\end{figure}

		\begin{figure*}[t]
			\begin{center}
				(a)\includegraphics[width=0.9\columnwidth,keepaspectratio,clip]{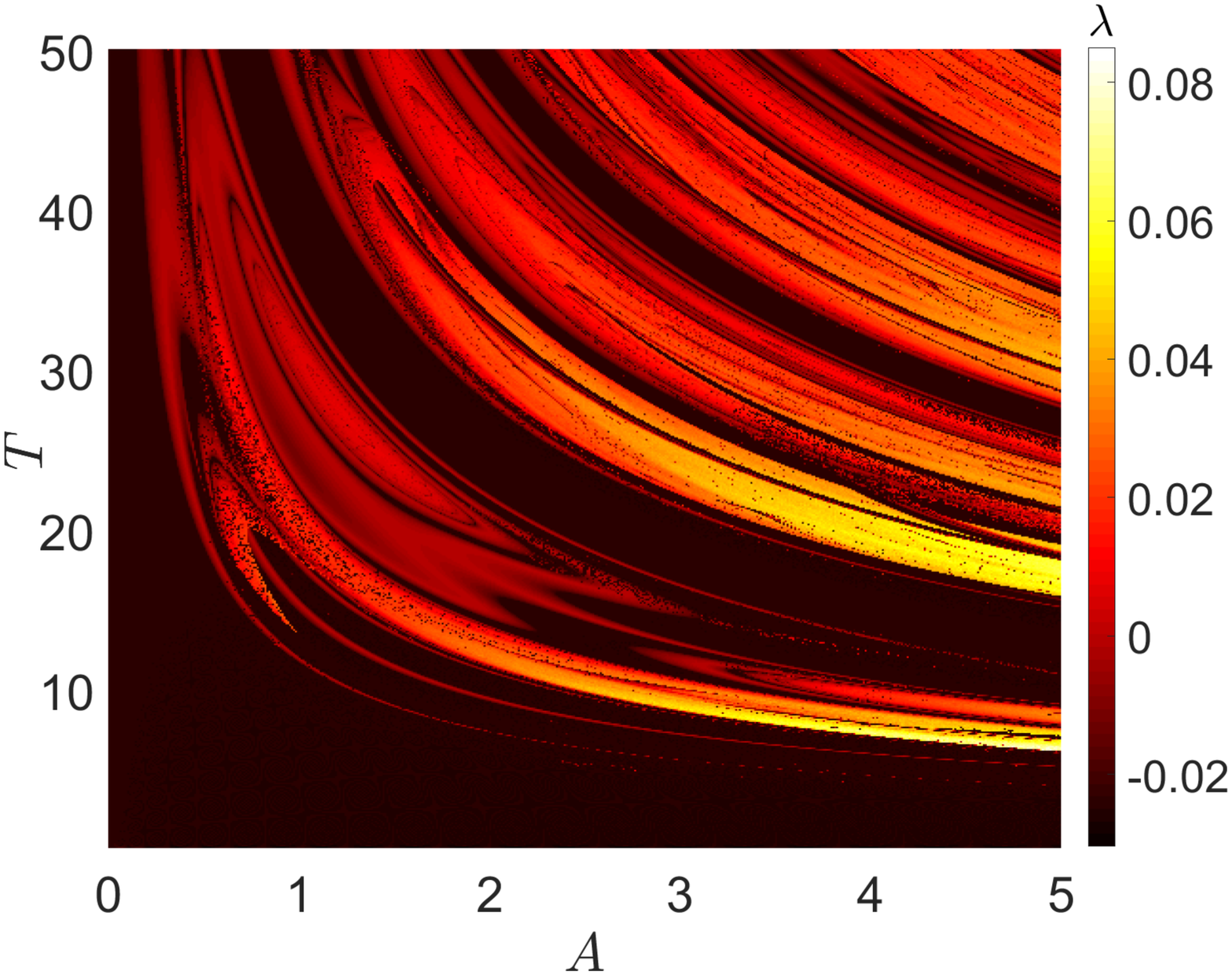}
				(b)\includegraphics[width=0.9\columnwidth,keepaspectratio,clip]{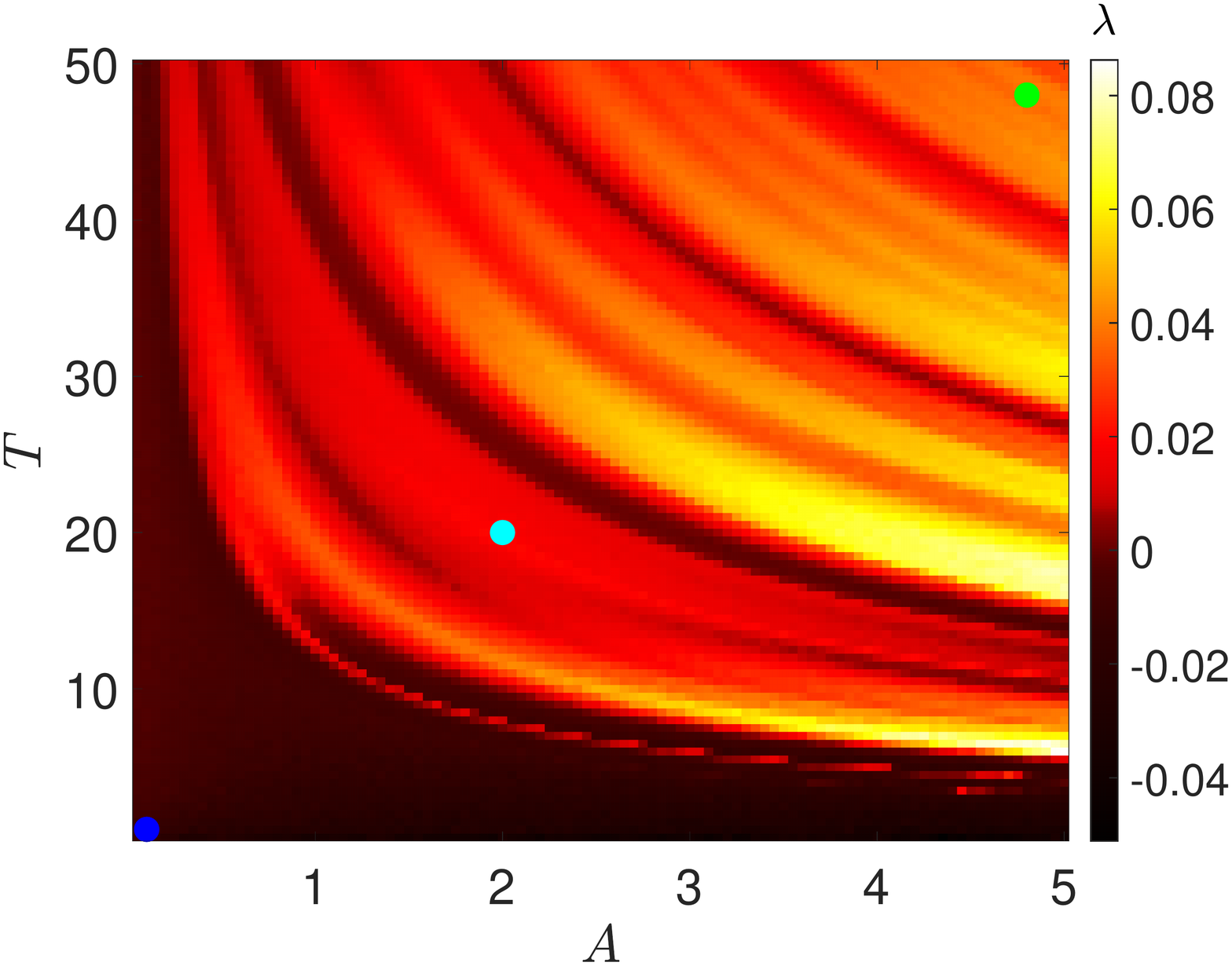}
				\caption{(Color online) Largest Lyapunov exponent as a function of amplitude $A$ and period $T$ of the modulations  
				for the model system in (a) the mean-field and (b) quantum versions (selected points correspond to those in Fig.\ref{fig:6}(b)).  Here $N=300$.}
				\label{fig:4}
			\end{center}
		\end{figure*}

		\section{Quantum Lyapunov exponent} To calculate the largest quantum LE, we employ the recently proposed method \cite{Yusipov2019}.
		It is based on the parallel evolution of fiducial and auxiliary trajectories, $\psi_f(t)$ and $\psi_a(t)$, under Eq.~(\ref{eq:4}), 
		in the spirit of the classical LE calculation \cite{Benettin1976}. The auxiliary trajectory is initialized 
		as a normalized perturbed vector $\psi_a^{init}=\psi_f^{init}+\varepsilon\psi_r$, produced with random i.i.d. entries in $\psi_r$ and $\varepsilon\ll1$. 
		The fiducial and perturbed observables, $\xi_f(t)$ and $\xi_a(t)$, 
		typically remain close to each other over many quantum jump events; 
		as the difference in the observables gets over an upper threshold, $\Delta(t_k)=|\xi_f(t_k)-\xi_a(t_k)|>\Delta_{max}$, or below a bottom threshold $\Delta(t_k)<\Delta_{min}$,
		the perturbed state is set back closer to the fiducial one along the mismatch direction $\psi_f(t_k)-\psi_a(t_k)$, 
		so that $|\xi_f(t_k)-\xi_a(t_k)|=\Delta_{0}$; the wave vector gets normalized and the occurred growth rate recorded, $d_k=\Delta(t_k)/\Delta_0$  \cite{periodic}. 
		The largest LE is estimated following the divergence of the chosen observable as $\lambda(t)=\frac{1}{t}\sum_k\ln d_k$ \cite{Benettin1976}.  
		
		We use  a recent high-performance realization 
		of the quantum jumps method \cite{Volokitin2017} to
		generate $M_{\mathrm{r}}=10^2$ different trajectories. First, we allow each trajectory to evolve up to $t_0 = 2\cdot10^3 T$ in order to arrive to the asymptotic regime, 
		and then we follow the dynamics of fiducial and auxiliary trajectories  up to $t = 10^3T$. We determine the dimension of the computational model, $N$, that would be large enough to ensure independence of the results on the chosen value. It has turned out that $N=150$ is already sufficient for the consistent calculation of the quantum LEs almost in the whole range of modulation amplitude, $A$ (cf. Fig.\ref{fig:3}(b), bottom panel). However, for $A>3.5$ oscillations of the number of photons in the cavity grow, and that leads to the distortion of the dynamics and positive LEs. Ultimately, we find it out that choosing the maximal number of photons $N>250$ provides consistency in the whole range of studied parameters.
		  
		Additionally, we verified convergence of the maximal LE to its asymptotic values in the regular and chaotic regimes and 
		confirm that the results are weakly dependent on the choice of observables (e.g., we used 
		photon population number, $n(t)=\langle\psi^\dagger(t)|\hat{n}|\psi(t)\rangle$ as an alternative); see Fig.~\ref{fig:2}.

		\begin{figure*}[t]
			\begin{center}
				(a)\includegraphics[width=0.9\columnwidth,keepaspectratio,clip]{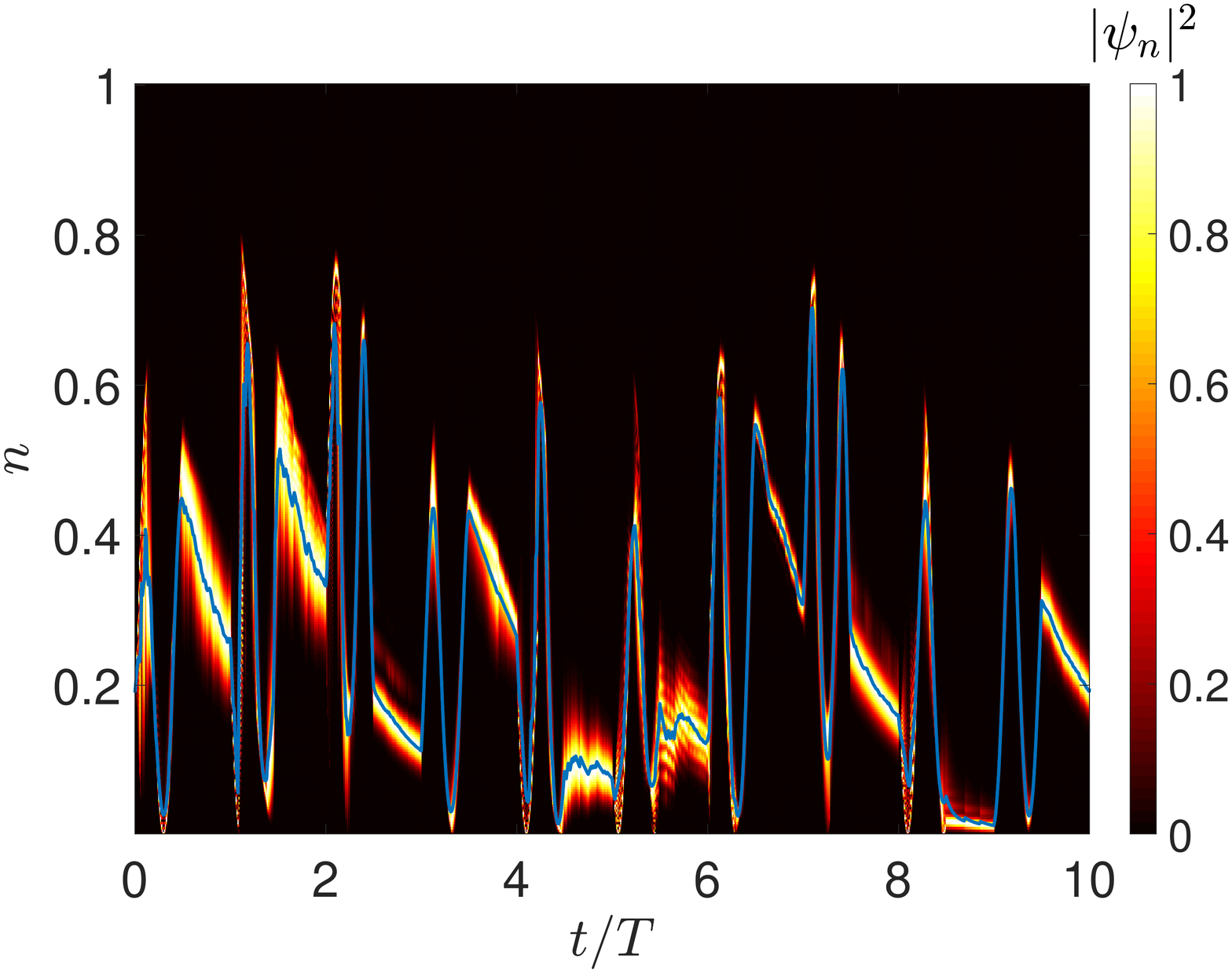}
				(b)\includegraphics[width=0.9\columnwidth,keepaspectratio,clip]{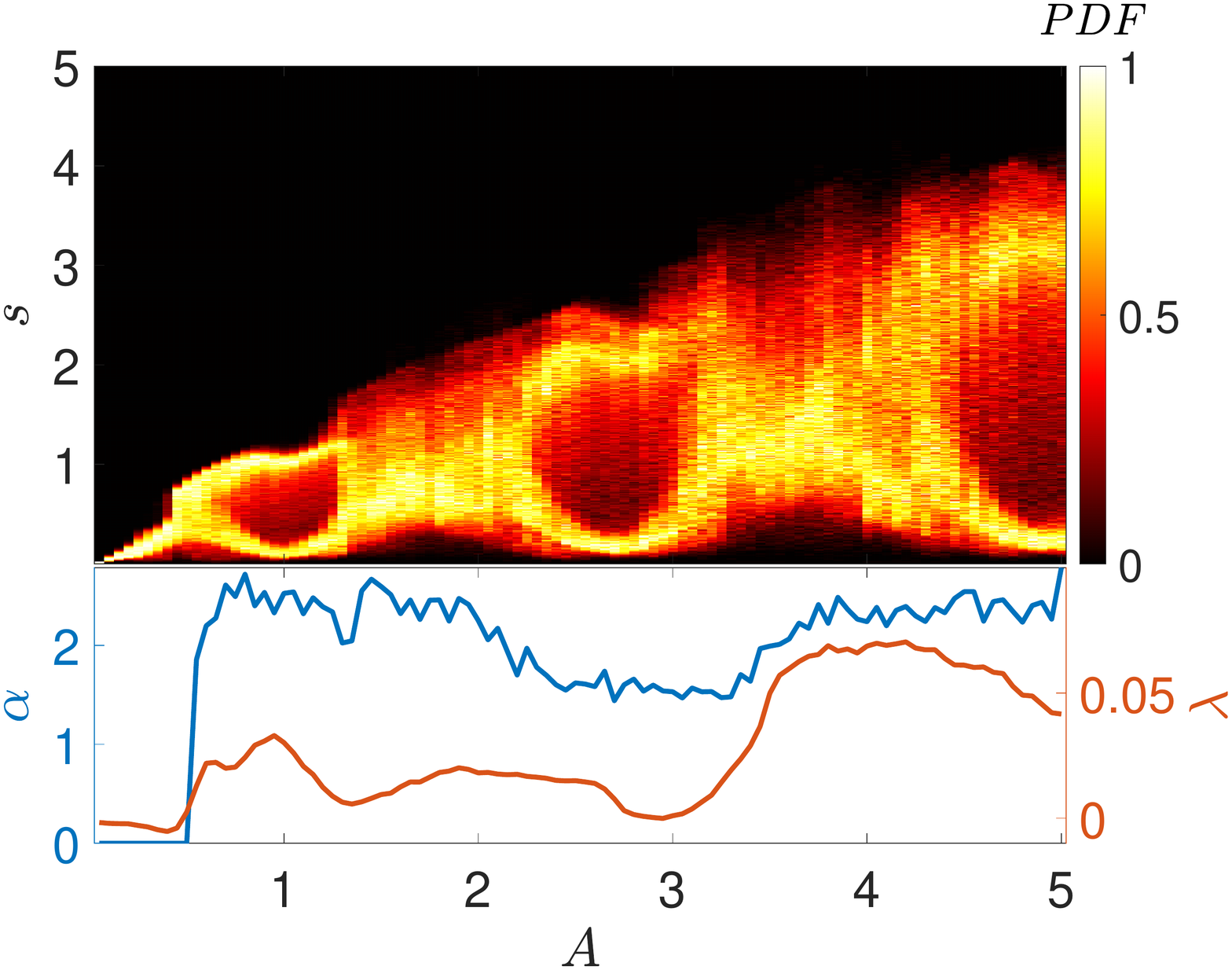}
				\caption{(Color online)  (a) Wave function evolution for an individual quantum trajectory deep in the chaotic regime for $A=4.0, T=20$ (color) and 
				expectation $n(t)$ (blue line). 
					(b) PDF of average norm decay rates, $W_s(s)$, between quantum jumps (top). Corresponding largest quantum LE (red) and an exponent for the power-law fit of $W_\tau(\tau)$ (blue, where valid). Note simultaneous broadening of $W_s(s)$, transition to  $\lambda>0$, and appearance of the power-law interval in the time between quantum jumps distribution.  Here $N=300$.} 
				\label{fig:5}
			\end{center}
		\end{figure*}

		Depending on  parameter values, interaction with the environment can strongly localize quantum trajectories 
		on the classical ones~\cite{Spiller1994,Brun1996,Jacobs2000,Pattanayak2018}. 
		Our case is notably different, so that the resulting structure of the probability distribution 
		for $\xi$ has only a qualitative resemblance, see Fig.~\ref{fig:3}(b), top. 
		Nevertheless, working in the essentially quantum regime, 
		we observe that the largest quantum LE becomes positive, see Fig.~\ref{fig:3}(b). Note, that the chaotic interval in the quantum case is somewhat greater, and fine structure intervals of classical regular dynamics are not reproduced. Such examples, when quantization generates chaotic dynamics, are known in the literature \cite{Pattanayak2018}.
		
		To get a more general picture, 
		we performed an extensive round of calculations and obtained  the two-parameter dependence of classical and quantum largest LEs on the amplitude and period of modulations; see Fig.~\ref{fig:4}(a,b). 
		Both the mean-field and quantum models produce visually similar structures of regular and chaotic regimes.

		\begin{figure*}[t]
			\begin{center}
				(a)\includegraphics[width=0.9\columnwidth,keepaspectratio,clip]{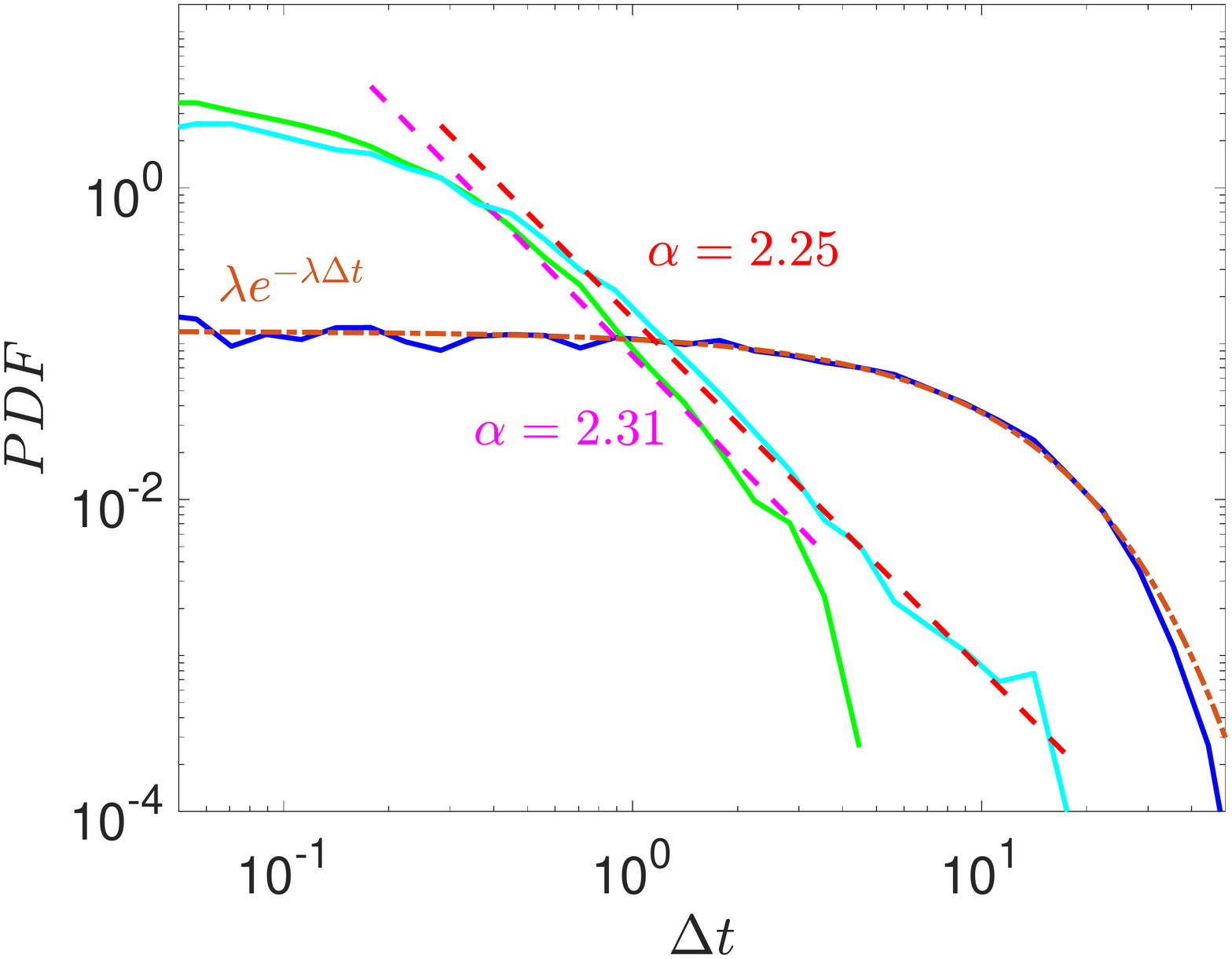}
				(b)\includegraphics[width=0.9\columnwidth,keepaspectratio,clip]{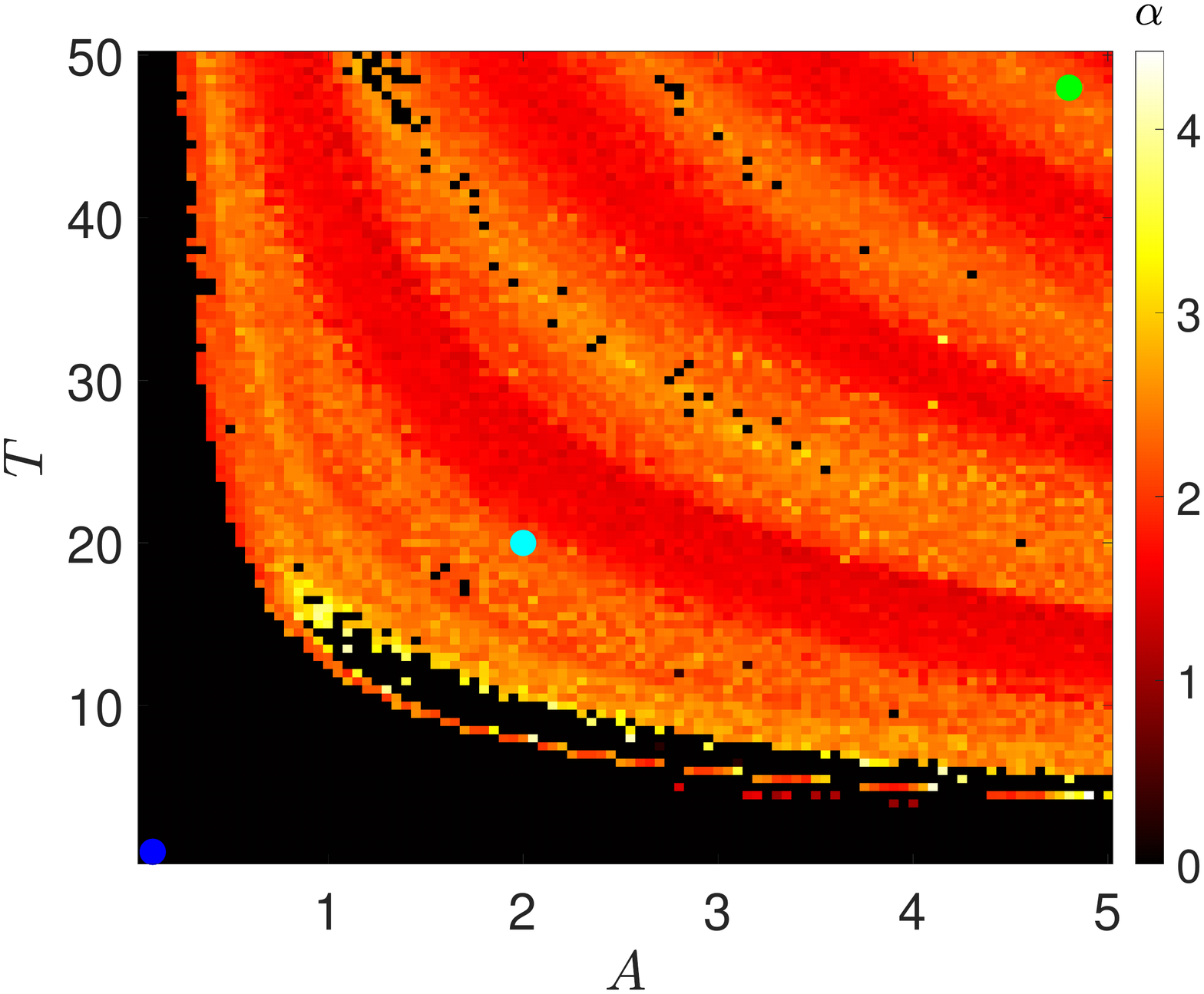}
				\caption{(Color online)  (a) PDF of time intervals between quantum jumps for selected parameter 
					values, $A=0.1, T=1$ (blue), $A=2.0, T=20$ (cyan), and $A=4.8, T=48$ (green), cf. color matched points in Fig.\ref{fig:4}(b) and Fig.\ref{fig:6}(b), 
					power-law and exponential fits. (b) the heat map for the power law exponent on the amplitude--period of modulation parameter plane, black color indicates its failure (compare to the largest LE diagram, Fig.\ref{fig:4}(b)). Here $N=300$.} 
				\label{fig:6}
			\end{center}
		\end{figure*}

		\section{Waiting time distribution} As we pointed it out in Introduction, quantum LEs are good objects for theoretical and numerical analysis but their experimental estimation is highly nontrivial. 
		
		Can we infer information about intra-cavity dynamics from what is accessible in experiment? Statistics of fluorescent photon emission from cavity is one of the most popular characteristics in quantum optics and appears promising for our purpose \cite{Carmichael1989,Delteil2014,Saez2018,Brange2019}.
		The other reason for the choice is that there is only one dissipative channel in our model system, and a single  quantum jump corresponds to a single photon emission, so that the intra-cavity dynamics and radiation are tightly bound.

		Evolution of the wave vector norm between consecutive  quantum jumps at times $\{t_k\}$ is governed by Eq.~(\ref{eq:4a}). For the system under study,  Eqs.(\ref{eq:2},\ref{eq:3}), it follows 
			\begin{equation}
				\label{eq:7a}
				\frac{d}{dt}||\psi||= -\gamma n(t)||\psi||,
				\end{equation}
				as $V^{\dagger}V=\gamma\hat{n}$. 
		The norm decay from $||\psi(t_{k-1})||=1$ to the random jump threshold $||\psi(t_{k})||=\eta_k$ 
		can be approximated by an exponent with an average rate proportional to an effective number of photons within $t_{k-1}<t<t_k$, that is $s_k=\gamma n^{(eff)}_{k}$. In other words, one can write $\tau_k=t_{k}-t_{k-1}=-\ln(\eta_k)/s_k$. 
		Since $\eta_k$ are random i.i.d. numbers on $[0,1]$, the variable $\zeta=-\ln(\eta_k)$ has the probability density distribution $W_\zeta(\zeta)=\exp(-\zeta)$. 
		If an asymptotic density matrix $\rho$ has a regular structure (e.g. unity matrix or unimodal distribution) then $n\approx const$ and, hence, $s\approx const$. In this case, the intervals between jumps also follow Poisson distribution, $W_\tau(\tau)=\exp(-\tau)$. Another example is a bimodal asymptotic solution, a result of the quantum analogue of period doubling bifurcation \cite{Ivanchenko2017,Poletti2017}, also a regular one, that would produce a superposition of two exponents.

		When the system has a chaotic attractor, the dynamics of observables, in particular, $n(t)$, becomes irregular. An example for a quantum trajectory deep in the chaotic regime, $A=4.0, T=20$, is presented in Fig.\ref{fig:5}(a). It is seen that the norm decay rates between each jump differ substantially. Therefore, the probability distribution $W_s(s)$ cannot be deduced from simple argumetns any longer. At the same time, it can be estimated numerically, see Fig.\ref{fig:5}(b), upper panel. The results indicate that the distribution becomes much broader for $A>0.4$, with transition to chaos (Fig.\ref{fig:5}(b), lower panel). Correspondingly, the resulting distribution of interjump intervals, $W_\tau(\tau)$, will no longer be exponential. In the complete absence of a theoretical background in this case, we turn to numerical simulations.


		Our main finding is presented in Fig.~\ref{fig:6}.
		The key observation is that the waiting time distribution  becomes  non-Poissonian and acquires a power-law intermediate scaling when 
		the largest quantum LE becomes positive, Fig.~\ref{fig:6}, see also Fig.\ref{fig:5}(b), lower panel.

		We performed massive sampling over the parameter plane $\{A,T\}$. 
		At every parameter point, the power-law fitting of the sampled PDF was obtained by the least squares linear regression for the log-log scaled distributions. The quality of fit was characterized by the coefficient of determination \cite{Regression}, $R^2\in[0,1]$, with larger values corresponding to better fit. 
	We vary the interval of durations to search for the best fit, the power-law exponent $\alpha$, and request that it spans over at least one decade for the horizontal axis and has $R^2>0.98$ \cite{note2}. Otherwise, the power-law hypothesis is rejected (black zones in Fig.~\ref{fig:6}(b)). 
		
		It is noteworthy that the zones, where power-law asymptotics in the waiting time distribution is present, are well correlated with the zones, where the corresponding maximal quantum LE is positive (cf. Fig.\ref{fig:4}(b)). Moreover, there are concordant variations of the values of power-law exponents and of the largest LE.

		\section{Conclusions}
		
		By using an open ac-driven Kerr cavity  as a model, we found that the photon waiting time statistics can serve a good diagnostic tool to detect  dissipative quantum chaos by the appearance of power-law intermediate asymptotics (which can be alternatively 
		characterized by the positive largest quantum LE). 
		The theoretical foundation of this observation remains a challenge. We conjecture that the first step would be to assume the statistical independence of the consecutive times $t_{j-1}, t_j, t_{j+1}$ 
		(which is not guaranteed though), and then to implement the  machinery used by Scherr and Montroll 
		to derive power-law statistics of photocurrent in amorphous materials \cite{Scher1975}. Implementation of this program is the subject of future study.

		Our results open a new perspective for quantification of regimes emerging in open quantum systems far from equilibrium, 
		especially in such fields as quantum electrodynamics, quantum optics, and polaritonic devices, 
		where photon emission statistics is an established and conventional  tool  \cite{Carmichael1989,Delteil2014,Saez2018,Brange2019}. 
		Recently, a substantial progress in a non-demolition detection of individual 
		emitted microwave photons has been achieved (see, e.g., \cite{nondem2019} ) so that the field of microwave quantum optics represents the most appealing opportunity to explore dissipative quantum chaos by waiting time statistics.
		
		Potential links to self-organized criticality \cite{Markovic2014,Munoz2018} and L\'{e}vy flights (recently used to	model  power-law  flip statistics 
		of open spin systems \cite{Imamoglu2016}) are another issues worth of consideration.        
		
		The authors acknowledge support of Basis Foundation 
		grant No. 17-12-279-1 and President of Russian Federation grant No. MD-6653.2018.2. Numerical simulations were carried out at the Lobachevsky University and Moscow State University supercomputers.


\begin{thebibliography}{99}
\bibitem{note2} We find that taking more or less stringent conditions for the validity of the power law fit can make a difference in `marginal' zones , where the largest LE is next to zero, and quantum chaos is weak or absent. These zones correspond to periodic windows in the classical system bifuration diagram. However, the main border between regularity and chaos is identified robustly.
\end{thebibliography}
	\end{document}